\documentclass[runningheads]{llncs}
\usepackage[T1]{fontenc}
\usepackage{graphicx,verbatim,multirow}
\usepackage{amsmath,amssymb}
\begin{document}
\title{A Surface-based Multimodal Framework for  Multitask Analysis in Alzheimer's Disease}
\titlerunning{Surface-based Multimodal AD Analysis Framework}
%

\author{Shuo Huang\inst{1,2,*} \and Jianwei Zhang\inst{1,3,*} \and
Lujia Zhong\inst{1,3} \and Jiaxin Yue\inst{1,3} \and Yihao Xia\inst{1} \and Xinkai Wang\inst{1,3} \and
Yonggang Shi\inst{1,2,3}}
%
\authorrunning{S. Huang et al.}
\institute{Stevens Neuroimaging and Informatics Institute, Keck School of Medicine, University of Southern California (USC), Los Angeles, CA 90033, USA \and
Alfred E. Mann Department of Biomedical Engineering, Viterbi School of
Engineering, University of Southern California (USC), Los Angeles, CA 90089, USA \and
Ming Hsieh Department of Electrical and Computer Engineering, Viterbi School of
Engineering, University of Southern California (USC), Los Angeles, CA 90089, USA\\
$^*$ S. Huang and J. Zhang contributed equally to this work.\\
\email{yshi@loni.usc.edu}}
  
\maketitle              
\begin{abstract}
Alzheimer's Disease (AD) is a progressive neurodegenerative disorder, and longitudinal analysis is critical for early detection and effective intervention. Developing models capable of multimodal and multitask analysis enables a more comprehensive understanding of AD progression. However, multimodal learning remains challenged by cross-modal misalignment, non-Euclidean surface representations of cortical data, and limited data availability in small-sample clinical settings. In this work, we propose an augmented spherical data-driven multimodal framework for multitask AD analysis. A spherical diffusion model is first trained to generate paired cortical thickness and Tau PET Standardized Uptake Value Ratio (SUVR) data, enabling structurally consistent multimodal augmentation on cortical surfaces while preserving anatomical correspondence. The augmented data are subsequently used to train a contrastive learning model that learns aligned and fused cross-modal representations. This design strengthens multimodal integration and encourages more balanced representation learning. The learned imaging features are further integrated with tabular cognitive assessments and demographic variables, and processed using an in-context learning model to perform both classification and regression tasks without task-specific fine-tuning. Experiments on the Alzheimer's Disease Neuroimaging Initiative (ADNI) dataset ($n = 802$) demonstrate consistent performance improvements across five diagnostic and longitudinal tasks, outperforming six baseline models.


\end{abstract}
\section{Introduction}
Alzheimer’s Disease (AD) is a progressive neurodegenerative disorder that evolves from cognitively normal (CN) to mild cognitive impairment (MCI) and eventually AD~\cite{shengraphmulti}. Early diagnosis and accurate modeling of disease progression are essential for timely intervention. However, heterogeneous disease trajectories make reliable prediction challenging.

Multimodal data, including medical images, neuroimaging biomarkers, cognitive assessments, and genetic information, provide complementary evidence for AD diagnosis and longitudinal analysis~\cite{multidiag,multimodalresults,multimodalcla,multiai,swinmultimodal,multimodacontr,multimodagui}. In particular, cortical thickness derived from T1-weighted MRI reflects neurodegeneration and cortical atrophy, while Tau PET Standardized Uptake Value Ratio (SUVR) maps on cortical surfaces capture pathological protein accumulation associated with disease progression. These two modalities characterize distinct yet biologically related aspects of AD and are therefore highly complementary. Furthermore, these  cortical measurements are naturally defined on curved surfaces rather than regular grids. Spherical representations of these modalities can preserve anatomical topology and inter-subject alignment~\cite{spheune,sphedeforunet}, but require specialized modeling for non-Euclidean structures. Meanwhile, multimodal learning remains challenged by modality misalignment~\cite{shenmulti} and limited data availability~\cite{synthedata}, which restrict generalization in small-sample settings. To alleviate the challenge of limited sample size, generative models such as VAEs, GANs, and diffusion models have been explored for neuroimaging data augmentation~\cite{vaegener,generimag,multimofusi}. However, existing studies largely focus on volumetric or single-modality generation, with limited investigation of paired multimodal surface synthesis and its impact on downstream multitask AD analysis.

To address these challenges, we propose in this work a surface-based  multimodal framework for multitask AD analysis. A spherical diffusion model is first trained to generate paired cortical thickness and Tau PET SUVR data, enabling structurally consistent multimodal augmentation. A spherical contrastive learning model is then used to learn aligned surface representations, which are integrated with demographic, genetic, and cognitive variables for classification and regression under an in-context learning framework. Experiments on the Alzheimer's Disease Neuroimaging Initiative (ADNI) dataset~\cite{adnioverview,adnidatause} ($n=802$) demonstrate consistent improvements across five tasks compared with six baseline models. Shapley additive explanations (SHAP) analysis~\cite{shapcalculate,shapcode} further quantifies modality contributions.

\begin{figure}[tb]
\includegraphics[width=0.99\textwidth]{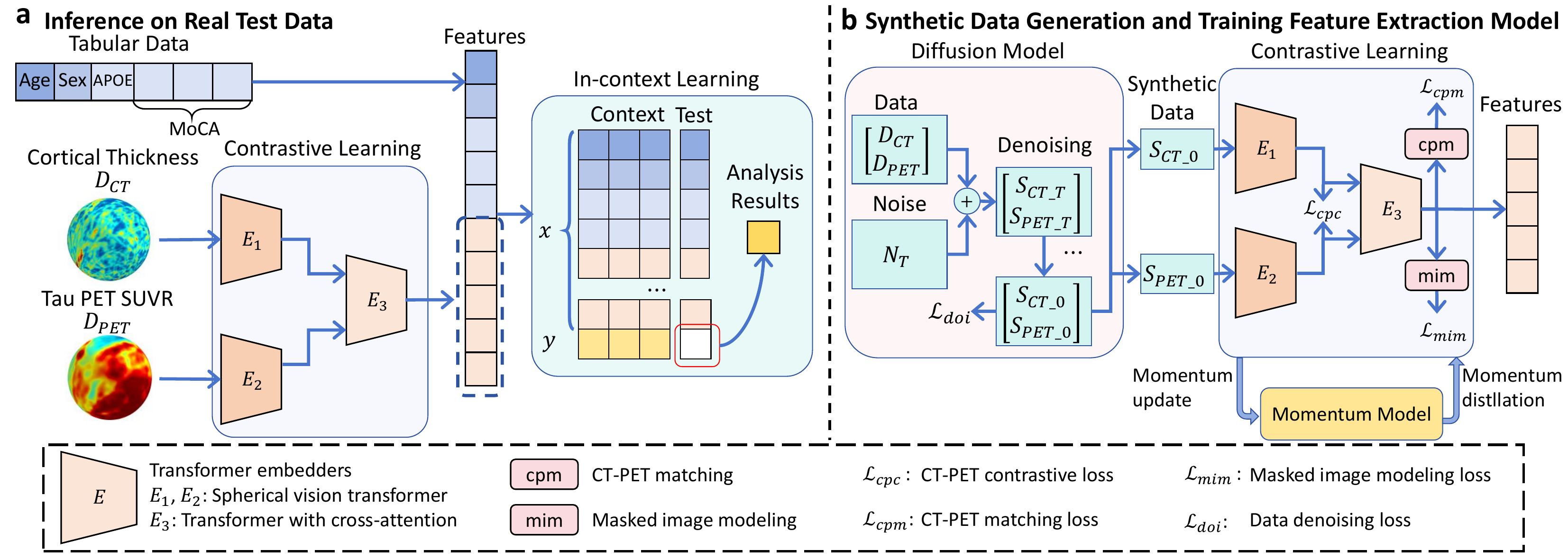}
\centering
\caption{Workflow of the proposed model. \textbf{(a)} Inference stage: spherical imaging data are processed by an unsupervised contrastive learning model to extract fused representations. The fused features are combined with tabular data and fed into an in-context learning framework for final prediction. \textbf{(b)} Training stage: the contrastive learning model is trained using augmented data generated by a spherical diffusion model.} \label{FigFuture_workflow}
\end{figure}

\section{Method}
Fig.~\ref{FigFuture_workflow} illustrates the overall framework of the proposed method. Our framework consists of surface-based data augmentation, cross-modal representation extraction, and in-context multimodal analysis.

\subsection{Spherical Diffusion for Data Augmentation}
To enrich the training data and enable paired multimodal data augmentation, we employed the spherical denoising diffusion probabilistic model (DDPM) \cite{zhang2025anatomy} 
to jointly generate cortical thickness (CT) and Tau PET SUVR (PET) data. 
For each modality, the left and right hemispheres were concatenated along the channel dimension, 
and the two modalities were further concatenated to form a unified input $D$. Following the standard DDPM formulation \cite{ho2020denoising}, 
Gaussian noise was added according to a predefined variance schedule, yielding

\begin{equation}
S_t = \sqrt{\bar{\alpha}_t}\, D + \sqrt{1 - \bar{\alpha}_t}\,\epsilon,
\end{equation}

\noindent where $S_t$ is the noisy data at time step $t$ and $D$ is the original data. 
$\epsilon \sim \mathcal{N}(0, I)$ is shared across all modalities to preserve cross-modality correspondence. 

The denoising network is implemented with self-attention layers operating on the joint channel representation, 
which enables modeling of cross-modality and cross-hemisphere dependencies during reverse diffusion. 
As a result, the generated samples maintain paired structural consistency between CT and PET.

The model was trained with a maximum diffusion step $T_{\text{train}} = 1000$, 
and during augmentation, reverse denoising was performed from selected diffusion steps to generate diverse paired samples. The augmented data were used to train the downstream model.


\begin{figure}[tb]
\centering
\includegraphics[width=0.94\textwidth]{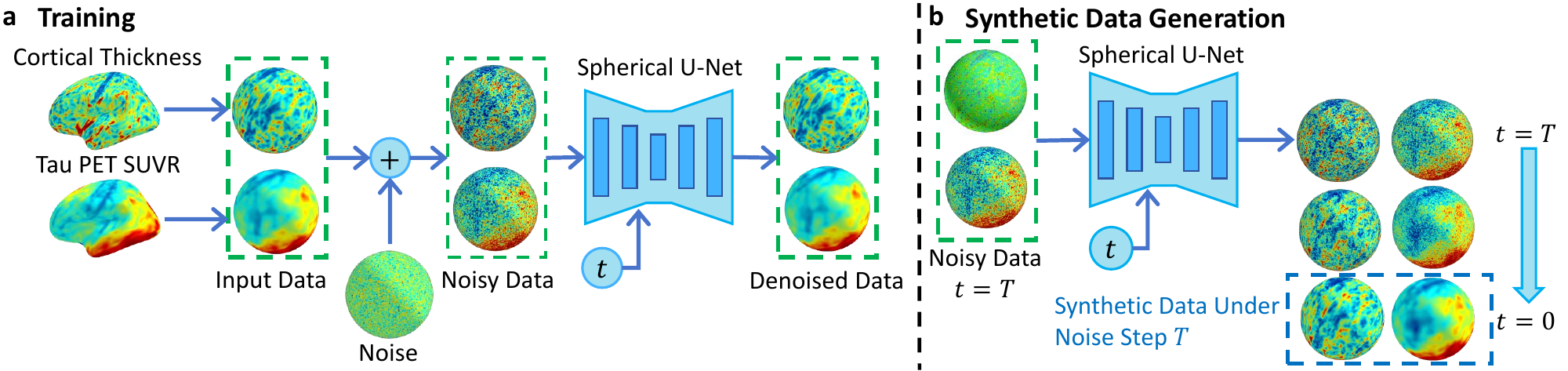}
\caption{Spherical diffusion model for paired CT and Tau PET surface data generation. \textbf{(a)} Training with shared noise across modalities. \textbf{(b)} Reverse denoising is initiated from a selected diffusion step $T$ and proceeds to $0$, producing structurally consistent paired surface samples.} \label{Figure_diffusion}
\end{figure}


\begin{figure}[tb]
\includegraphics[width=0.75\textwidth]{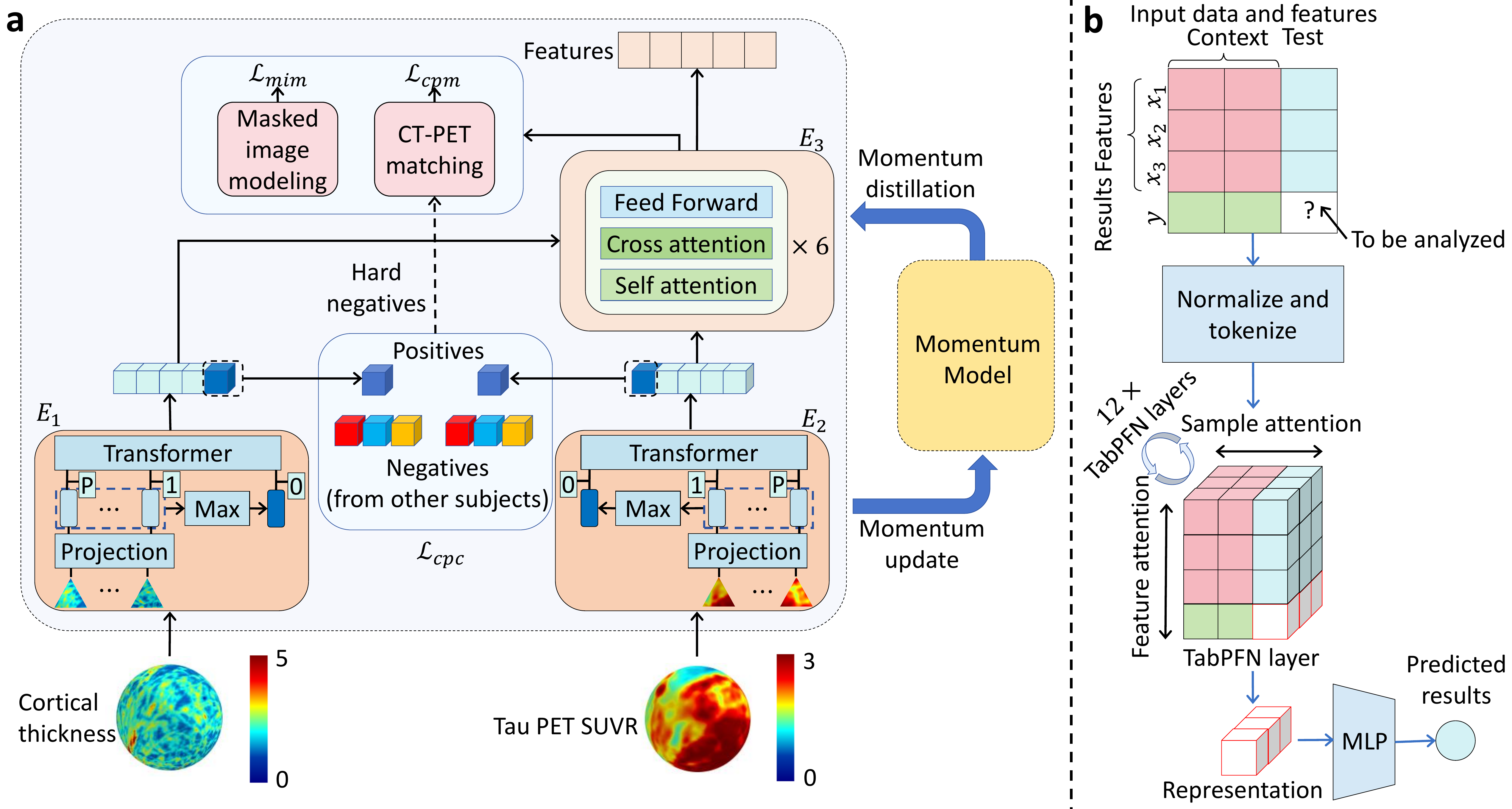}
\centering
\caption{The contrastive multimodal fusion using the SViT embedding method (\textbf{a}) and the test workflow for one data using in-context learning by TabPFN (\textbf{b}).}
\label{FigFuture_contrast}
\end{figure}

\subsection{Cross-modal Alignment Trained by Augmented Data}

Based on the augmented surface data, we learn aligned and fused multimodal representations through contrastive learning. Following ViT \cite{dosovitskiy2020image,dahan2022surface} and ALBEF~\cite{alignbefor}, 
we employed a spherical vision transformer (SViT) and a contrastive learning framework, as shown in Fig. \ref{FigFuture_contrast}a.

\textbf{SViT.} Cortical surfaces are resampled onto an $i$-th order icosahedral mesh, where coarse triangular faces at order $(i-j)$ are mapped onto the fine mesh and partitioned into geodesic patches via shortest-path boundary construction. 

Given input $\mathbf{X}$, 
each patch $\mathbf{x}_p$ is projected by a learnable matrix $W$ to obtain token embeddings $\mathbf{e}_p = W\mathbf{x}_p$.
An initial class token $\mathbf{e}_{\text{cls}}$ is constructed via max pooling, and the sequence $\mathbf{E}_0=[\mathbf{e}_{\text{cls}},\mathbf{e}_1,\dots,\mathbf{e}_P]$
is processed by a Transformer encoder, producing output tokens $\mathbf{Z}=[\mathbf{z}_{\text{cls}},\mathbf{z}_1,\dots,\mathbf{z}_P]$. The refined class token $\mathbf{z}_{\text{cls}}$ serves as the global surface representation.

\textbf{Modality Encoding.} CT and PET are independently encoded by SViT, yielding $\mathbf{Z}^{CT}=[\mathbf{z}^{CT}_{\text{cls}},\mathbf{z}^{CT}_1,\dots,\mathbf{z}^{CT}_P]$, and $\mathbf{Z}^{PET}=[\mathbf{z}^{PET}_{\text{cls}},\mathbf{z}^{PET}_1,\dots,\mathbf{z}^{PET}_P]$.
The alignment is enforced via the contrastive loss

\begin{equation}
\mathcal{L}_{CPC}=
-\log 
\frac{\exp(\mathrm{sim}(\mathbf{z}^{CT}_{\text{cls}},\mathbf{z}^{PET}_{\text{cls}})/\tau)}
{\sum_{k}\exp(\mathrm{sim}(\mathbf{z}^{CT}_{\text{cls}},\mathbf{z}^{PET}_{\text{cls},k})/\tau)},
\label{eq:CPC}
\end{equation}

\noindent where $\mathrm{sim}(\cdot,\cdot)$ denotes cosine similarity, 
$\tau$ is a temperature parameter, and $k$ indexes negative samples.

\textbf{Cross-Modal Fusion.} Cross-attention operates \cite{vaswani2017attention} on the encoded tokens and produces fused representations $\mathbf{Z}^{fus}=[\mathbf{z}^{fus}_{\text{cls}},\mathbf{z}^{fus}_1,\dots,\mathbf{z}^{fus}_P]$.
A matching loss is defined using cross-entropy loss $\mathrm{CE}(\cdot)$ as

\begin{equation}
\mathcal{L}_{CPM}=\mathrm{CE}(f(\mathbf{z}^{fus}_{\text{cls}}),y_{match}),
\label{eq:CPM}
\end{equation}

\noindent where $f(\cdot)$ is a classifier and $y_{match}$ is the binary CT-PET matching label.

\textbf{Masked Image Modeling.} Masked modeling is performed on the Transformer output tokens of the PET branch. 
A subset of indices $\mathcal{M}$ is randomly selected from $\{\mathbf{z}^{PET}_p\}_{p=1}^P$, and the model reconstructs the masked encoded tokens by minimizing

\begin{equation}
\mathcal{L}_{MIM}=
\frac{1}{|\mathcal{M}|}
\sum_{i\in\mathcal{M}}
\left\|
\mathbf{z}^{PET,\,orig}_i -
\mathbf{z}^{PET,\,reco}_i
\right\|_2^2,
\label{eq:MIM}
\end{equation}

\noindent where $\mathbf{z}^{PET,\,orig}_i$ denotes the original encoded token and 
$\mathbf{z}^{PET,\,reco}_i$ denotes its reconstructed counterpart.

\textbf{Momentum Stabilization.} A momentum encoder is updated via exponential moving average of the original model to provide stable training targets. 
The overall objective is 
$\mathcal{L}=(1-\alpha)\mathcal{L}_{ori}+\alpha\mathcal{L}_{KL}$, 
where $\mathcal{L}_{ori}=\mathcal{L}_{CPC}+\mathcal{L}_{CPM}+\mathcal{L}_{MIM}$, 
$\mathcal{L}_{KL}$ is a regularization term, and $\alpha=0.4$ balances the two components.

\subsection{In-context Multimodal Analysis}

To integrate imaging and tabular information for downstream prediction, we adopt an in-context learning strategy based on TabPFN~\cite{incontexttable,exptabincont,tabpmee}. Rather than optimizing model parameters on a fixed training set, TabPFN conditions each test prediction on the training data at inference time, as shown in Fig. \ref{FigFuture_contrast}b.



Let $\mathcal{D}_{train}=\{(x_i,y_i)\}_{i=1}^{n}$ denote the training set and $x_{test}$ a test sample. TabPFN estimates $p(y_{test}\mid x_{test}, \mathcal{D}_{train})$ by jointly processing $\mathcal{T}=\{(x_1,y_1),\dots,$
$(x_n,y_n),(x_{test},?)\}$. The concatenated table is normalized and tokenized before being passed through transformer layers that model inter-sample and inter-feature dependencies~\cite{transformercanlearn}. The representation corresponding to $x_{test}$ is then fed into a lightweight MLP head to produce the prediction. Pre-trained on large-scale synthetic tabular tasks, TabPFN approximates Bayesian posterior inference and generalizes effectively to small datasets without task-specific fine-tuning.



\subsection{Subject-level Representation Diversity}
Synthetic augmentation increases the number of training samples but does not necessarily increase representation diversity across subjects. To quantify this effect, we analyze subject-level diversity in the embedding space.

Given representations $z_i$ with subject labels $y_i$, we compute for each sample its strongest same-subject similarity $s_i^+ = \max_{j:y_j=y_i,\,j\neq i} \cos(z_i,z_j)$
and its strongest different-subject similarity (hard negative) $s_i^- = \max_{k:y_k\neq y_i} \cos(z_i,z_k)$.
We define the hard-negative score $h = \frac{1}{N}\sum_{i=1}^N \mathbf{1}(s_i^+ > s_i^-)$. Values of $h$ close to $1$ or $0$ indicate reduced representation diversity, whereas $h \approx 0.5$ reflects higher diversity.

Let $A$ denote task performance (e.g., F1). To jointly assess performance and representation diversity, we define the utility-diversity score (UDS) as
\begin{equation}
\text{UDS} = \sqrt{A \cdot (1 - 2|h - 0.5|)}.
\end{equation}

UDS is used to select the diffusion noise level that balances predictive accuracy and subject-level representation diversity.

\section{Results}
Experiments were conducted on the ADNI dataset~\cite{adnioverview,adnidatause} ($n=802$: 452 CN, 269 MCI, 81 AD; 393 males, $75.54\pm7.76$ years; 409 females, $72.70\pm8.22$ years). The data were randomly split into 480 training and 322 testing subjects. All experiments were performed on NVIDIA A6000 workstations.

We evaluated our method on both AD diagnosis and longitudinal prediction tasks. 
For classification, we report accuracy (Acc.), area under the ROC curve (AUC), 
macro-F1 (F1), macro precision (Prec.), macro recall (Recall), and Matthews correlation coefficient (MCC). 
For regression, we report $R^2$, root mean squared error (RMSE), mean absolute error (MAE), 
and Pearson and Spearman correlations. 
SHAP score \cite{shapcalculate,shapcode} were used to analyze modality contributions.


\begin{table}[t]
\centering
\caption{Noise step selection for spherical data. \textbf{Best value}; \underline{second-best value}.}
\label{Table_noise}
\begin{tabular}{|c| c| c| c| c| c| c| c| c| c| c|}
\hline
Method & Acc & Prec & Recall & F1 & AUC & MCC & HNIS & UDS \\
\hline
TabPFN (Ori. Data) & 0.5969 & 0.5645 & 0.4256 & 0.4305 & \underline{0.6876} & 0.2000 & -- & -- \\
Ours ($T = 50$)   & 0.5736 & \underline{0.5777} & 0.4847 & 0.5111 & 0.6660 & 0.1856 & 0.9719 & 0.1694 \\
Ours ($T = 100$)  & 0.5581 & 0.4772 & 0.4187 & 0.4281 & 0.6846 & 0.1568 & 0.9979 & 0.0424 \\
Ours ($T = 200$)  & \underline{0.6124} & \textbf{0.6068} & \underline{0.5477} & \textbf{0.5701} & \textbf{0.7187} & \underline{0.2798} & 0.0141 & 0.1269 \\
Ours ($T = 300$)  & \textbf{0.6202} & 0.5614 & \textbf{0.5964} & \underline{0.5684} & 0.6752 & \textbf{0.3106} & \textbf{0.4151} & \textbf{0.6869} \\
Ours ($T = 400$)  & 0.5659 & 0.5690 & 0.4769 & 0.5024 & 0.6660 & 0.1680 & 0.2021 & 0.4506 \\
Ours ($T = 500$)  & 0.5116 & 0.4218 & 0.3933 & 0.3972 & 0.6319 & 0.0607 & \underline{0.3911} & \underline{0.5574} \\
\hline
\end{tabular}
\end{table}

\subsection{Choosing the Best Noise Level}
\label{best_noise_step}
For data augmentation, the diffusion model was trained for 1000 epochs. Noise levels $T \in [50,500]$ were evaluated, and three denoised samples were generated per subject to construct the synthetic training set. A contrastive model was trained separately for each $T$ using only synthetic data, while TabPFN was applied exclusively to the original test data to avoid leakage.

The optimal noise level was selected using the UDS criterion based on macro-F1 for three-class CN/MCI/AD classification (Table~\ref{Table_noise}). Among the 322 test subjects, 193 were used as context samples and 129 for evaluation. Performance shows a non-monotonic trend: moderate noise improves results, whereas insufficient or excessive noise degrades representation quality.

As $T$ increases, HNIS approaches $0.5$, indicating greater representation diversity. At $T=200$, the near-zero HNIS reflects a bias toward hard-negative similarity, reducing diversity and UDS. At $T=300$, HNIS remains close to $0.5$ while predictive performance is maintained, yielding the highest UDS. We therefore select $T=300$ for subsequent experiments.

\begin{table}[t]
\centering
\caption{Multimodal performance comparison for three-class classification.}
\label{Table_multiclass}
\begin{tabular}{|c| c| c| c| c| c| c|}
\hline
Method & Acc & Prec & Recall & F1 & AUC & MCC\\
\hline
XGBoost \cite{xgboost} & 0.6279 & 0.6204 & 0.5978 & 0.5997 & 0.7259 & 0.3035 \\
Random Forest \cite{breiman2001random} & 0.5581 & 0.5849 & 0.4481 & 0.4767 & 0.6634 & 0.1465 \\
LightGBM \cite{lightgbm} & 0.5969 & 0.5695 & 0.5731 & 0.5648 & 0.7070 & 0.2486 \\
TabNet \cite{tabnet} & 0.5271 & 0.2909 & 0.3328 & 0.2981 & 0.5667 & 0.0110 \\
AutoGluon \cite{erickson2020autogluon} & 0.6202 & 0.6080 & 0.5153 & 0.5405 & 0.7352 & 0.2755 \\
MulT \cite{tsai2019multimodal} & 0.5969 & 0.5487 & 0.5954 & 0.5663 & 0.7526 & 0.2905\\
\hline
TabPFN (Tab. only) & 0.5814 & \underline{0.6830} & 0.5020 & 0.5464 & 0.7537 & 0.2141\\
TabPFN (Ori. Data) & \underline{0.6589} & 0.6424 & \underline{0.6013} & \underline{0.6113} & \underline{0.7863} & \underline{0.3618} \\
\hline
Ours ($T = 300$) & \textbf{0.6744} & \textbf{0.6902} & \textbf{0.6622} & \textbf{0.6709} & \textbf{0.8356} & \textbf{0.3982} \\
\hline
\end{tabular}
\end{table}


\begin{table}[t]
\centering
\caption{Performance comparison for 1-year and 2-year progressing data ($\Delta MoCA < 0$) detection in MCI and AD subjects.}
\label{Table_progress}
\begin{tabular}{|c| c| c| c| c| c| c| c| c|}
\hline
Method & Surf. & Tab. & Acc & Prec & Recall & F1 & AUC & MCC \\
\hline
\multicolumn{9}{|c|}{1 year} \\
\hline
TabPFN &  & \checkmark & 0.5806 & 0.5855 & 0.5833 & 0.5827 & 0.6125 & 0.1688 \\
TabPFN (Ori. Data) & \checkmark & \checkmark & 0.6129 & 0.6228 & 0.6167 & 0.6195 & 0.6300 & 0.2394 \\
Ours ($T = 300$) & \checkmark &  & 0.6129 & 0.6125 & 0.6125 & 0.6125 & 0.6292 & 0.2250 \\
Ours ($T = 300$) & \checkmark & \checkmark & \textbf{0.6452} & \textbf{0.6517} & \textbf{0.6479} & \textbf{0.6437} & \textbf{0.6417} & \textbf{0.2996} \\
\hline
\multicolumn{9}{|c|}{ 2 years} \\
\hline
TabPFN (Ori. Data) & \checkmark & \checkmark & 0.6429 & 0.6458 & 0.6429 & 0.6410 & 0.5306 & 0.2887 \\
Ours ($T = 300$) & \checkmark & \checkmark & \textbf{0.7143} & \textbf{0.7333} & \textbf{0.7143} & \textbf{0.7083} & \textbf{0.7143} & \textbf{0.4472} \\
\hline
\end{tabular}
\end{table}



\subsection{Analyzing AD using Multimodal Data}
Spherical imaging data (cortical thickness and Tau PET SUVR), demographic variables (age and gender), APOE genotype, and Montreal Cognitive Assessment (MoCA) scores were integrated for multimodal analysis.

\begin{table}[tb]
\centering
\caption{Longitudinal MoCA score prediction performance at 1-year and 2-year follow-up in MCI and AD subjects. $^{ns}$: $p \geq 0.05$, $^{*}$: $0.005 \leq p < 0.05$, $^{**}$: $p < 0.005$}
\label{Table_mocapredi}
\begin{tabular}{|c| c| c| c| c| c|}
\hline
Method & $R^2$ & RMSE & MAE & Pearson & Spearman \\
\hline
\multicolumn{6}{|c|}{1 year} \\
\hline
XGBoost \cite{xgboost} & 0.3873 & 4.3637 & 3.5586 & 0.6485$^{**}$ & 0.4862$^{*}$ \\
LightGBM \cite{lightgbm} & 0.2116 & 4.9500 & 3.6163 & 0.4633$^{*}$ & 0.4404$^{ns}$ \\
Random Forest \cite{breiman2001random} & 0.4148 & 4.2645 & 3.1897 & 0.7220$^{**}$ & 0.5011$^{*}$ \\
TabNet \cite{tabnet} & -14.82 & 22.18 & 21.42 & 0.0049$^{ns}$ & -0.0176$^{ns}$ \\
AutoGluon \cite{erickson2020autogluon} & 0.2823 & 4.7227 & 3.4885 & 0.6086$^{*}$ & 0.4738$^{*}$ \\
MuIT \cite{tsai2019multimodal} & 0.2808 & 4.7276 & 3.3312 & 0.5825$^{*}$ & 0.6092$^{*}$ \\
TabPFN (Ori. Data) & \underline{0.5216} & \underline{3.8558} & \underline{2.9387} & \underline{0.7350}$^{**}$ & \underline{0.7190}$^{**}$ \\
Ours ($T = 300$)   & \textbf{0.5556} & \textbf{3.7164} & \textbf{2.9251} & \textbf{0.7554}$^{**}$ & \textbf{0.7303}$^{**}$ \\
\hline
\multicolumn{6}{|c|}{2 years} \\
\hline
TabPFN (Ori. Data) & \underline{0.7409} & \underline{3.3421} & \underline{2.5098} & \underline{0.8792}$^{**}$ & \underline{0.8787}$^{**}$ \\
Ours ($T = 300$)   & \textbf{0.7835} & \textbf{3.0548}  & \textbf{2.1423} & \textbf{0.9103}$^{**}$ & \textbf{0.8954}$^{**}$ \\
\hline
\end{tabular}
\end{table}

We first evaluate the model on three-class CN/MCI/AD diagnosis, 
using 193 subjects as the context set and 129 for evaluation, 
consistent with Sec.~\ref{best_noise_step}. 
Longitudinal performance is then assessed on MCI and AD subjects from the test set. Four longitudinal tasks are considered: 
(i) progressing data detection, defined as $\Delta \text{MoCA} = \text{MoCA}_{\text{follow-up}} - \text{MoCA}_{\text{baseline}} < 0$,
and (ii) MoCA score prediction at 1-year and 2-year follow-ups. 
The 1-year and 2-year cohorts include 76 and 35 subjects, respectively.

Tables~\ref{Table_multiclass}-\ref{Table_mocapredi} summarize the results. 
Our method outperforms all comparison models across tasks. 
Ablation studies further demonstrate that multimodal input consistently surpasses single-modality tests. 
Performance on the 2-year tasks exceeds that of the 1-year tasks, suggesting that short-term (1-year) MoCA changes are less stable.

\subsection{Modality Contribution Analysis}
We conduct SHAP-based analysis to quantify modality contributions (Table~\ref{Table_shapleyanaly}). Overall, MoCA scores and spherical CT+PET features dominate model predictions, while demographic variables contribute minimally.

For three-class diagnosis, MoCA is more influential in distinguishing early stages, whereas imaging features contribute more prominently to AD identification. In 1-year progression detection, CT+PET features exhibit greater importance than MoCA, indicating that imaging features contribute more strongly to short-term progressing data detection. For longitudinal MoCA prediction, baseline MoCA remains the primary contributor.

These findings indicate that the proposed framework effectively leverages the complementary strengths of cognitive assessments and imaging biomarkers across tasks rather than relying on a single dominant modality.

\begin{table}[t]
\centering
\caption{SHAP-based modality contributions across tasks.}
\label{Table_shapleyanaly}
\begin{tabular}{|c| c| c| c| c|}
\hline
Task & CT+PET & MOCA & Gender & Others \\
\hline
3-class CN & \underline{0.1366 $\pm$ 0.0963} 
   & \textbf{0.2471 $\pm$ 0.1628} 
   & 0.0009 $\pm$ 0.0007 
   & 0.0011 $\pm$ 0.0008 \\
3-class MCI & \underline{0.2315 $\pm$ 0.2110} & \textbf{0.2392 $\pm$ 0.1616} & 0.0013 $\pm$ 0.0009 & 0.0011 $\pm$ 0.0008 \\
3-class AD & \textbf{0.0966 $\pm$ 0.2188} & \underline{0.0368 $\pm$ 0.0831} & 0.0004 $\pm$ 0.0007 & 0.0002 $\pm$ 0.0003 \\
\hline
Prog. 1 year & \textbf{0.2226 $\pm$ 0.1732} & \underline{0.0220 $\pm$ 0.0135} & 0.0001 $\pm$ 0.0001 & 0.0005 $\pm$ 0.0003 \\
Prog. 2 years & \underline{0.1458 $\pm$ 0.1070} & \textbf{0.1653 $\pm$ 0.1020} & 0.0038 $\pm$ 0.0007 & 0.0003 $\pm$ 0.0001 \\
\hline
MoCA 1 year & \underline{0.2730 $\pm$ 0.2084} & \textbf{2.3264 $\pm$ 2.4935} & 0.0126 $\pm$ 0.0089 & 0.0089 $\pm$ 0.0044 \\
MoCA 2 years & \underline{1.2070 $\pm$ 1.1669} & \textbf{2.0455 $\pm$ 1.1400} & 0.1977 $\pm$ 0.0684 & 0.0309 $\pm$ 0.0297 \\
\hline
\end{tabular}
\end{table}


\section{Conclusion}
In this work, we propose a surface-based multimodal framework for multitask AD analysis. By leveraging a spherical diffusion model to generate paired cortical thickness and Tau PET SUVR data, the proposed approach enables structurally consistent multimodal augmentation on cortical surfaces. The augmented data are used to train a contrastive learning model for cross-modal alignment, and the resulting imaging representations are integrated with clinical variables through an in-context learning framework for both classification and regression tasks. Experimental results on the ADNI dataset ($n=802$) show improvements across diagnostic and longitudinal tasks ($n=5$) on comparisons with baselines ($n=6$).

Future work will explore validation on larger and more diverse cohorts, investigate more advanced surface generative models, and extend the framework to additional neurodegenerative and clinical prediction tasks.

\begin{credits}
\subsubsection{\ackname} This work was supported by the National Institute of Health (NIH) under grants R01EB022744, RF1AG077578,  RF1AG064584,  U19AG078109, and P30AG066530. Authors thank the ADNI investigators (\url{https://adni.loni.usc.edu}).

\subsubsection{\discintname}
The authors have no competing interests to declare that are
relevant to the content of this article.
\end{credits}

\bibliographystyle{splncs04}
\bibliography{ref}
%




\end{document}